\documentclass[conference,a4paper]{IEEEtran}
\IEEEoverridecommandlockouts

\usepackage[hidelinks]{hyperref}
\usepackage[cmex10]{amsmath}
\usepackage{amssymb,amsfonts}
\usepackage{dblfloatfix}

\usepackage[ruled,vlined]{algorithm2e}
\usepackage{graphicx}
\usepackage{subcaption}

\usepackage{booktabs}
\usepackage{siunitx}
\usepackage[numbers,compress]{natbib}
\usepackage{texnames}
\usepackage{bm,bbm}
\usepackage{orcidlink}

\usepackage{float}

\usepackage{tikz}
\usetikzlibrary{shapes.geometric, arrows}

\tikzstyle{block} = [rectangle, draw, fill={rgb,255:red,23;green,190;blue,207}, fill opacity=0.3, text opacity=1, text width=12em, text centered, rounded corners, minimum height=2em]
\tikzstyle{doubleblock} = [rectangle, draw, fill={rgb,255:red,23;green,190;blue,207}, fill opacity=0.3, text opacity=1, text width=12em, text centered, rounded corners, minimum height=3em]
\tikzstyle{arrow} = [thick, ->, >=stealth]

\begin{document}

\title{\uppercase{To center or not to center? Hyperspectral data vs. quantum covariance matrices}
\thanks{AM was supported by the Priority Research Areas Anthropocene and Digiworld under the program Excellence Initiative – Research University at the Jagiellonian University in Kraków. JN and AMW were supported by the Silesian Univ. of Technology funds through the grant for maintaining and developing research potential.  }
}

\author{	\IEEEauthorblockN{Artur\ Miroszewski\orcidlink{0000-0002-0391-8445}}
	\IEEEauthorblockA{\textit{Institute of Theoretical Physics},\\
        \textit{Mark Kac Center for Complex}\\ \textit{Systems Research}\\
		Jagiellonian University\\
        Łojasiewicza 11, 30-348 Cracow, Poland\\
		artur.miroszewski@uj.edu.pl}
	\and
        \IEEEauthorblockN{Jakub\ Nalepa\orcidlink{0000-0002-4026-1569}}
	\IEEEauthorblockA{\textit{Silesian University of Technology} \\ Akademicka 2A, 44-100 Gliwice\\\textit{KP Labs}\\
        Bojkowska 37J, 44-100 Gliwice,\\Poland\\
		jnalepa@ieee.org}
        \and
	\IEEEauthorblockN{Agata\ M. Wijata\orcidlink{0000-0001-6180-9979}}
	\IEEEauthorblockA{\textit{Silesian University of Technology} \\ Akademicka 2A, 44-100 Gliwice\\\textit{KP Labs}\\
        Bojkowska 37J, 44-100 Gliwice,\\Poland\\
		awijata@ieee.org}
}

\maketitle
\begin{abstract}
    We highlight how the $L^2$ normalization required for embedding data in quantum states affects data centering, which can significantly influence quantum amplitude-encoded covariance matrices in quantum data analysis algorithms. We examine the spectra and eigenvectors of quantum covariance matrices derived from hyperspectral data under various centering scenarios. Surprisingly, our findings reveal that classification performance in problems reduced by principal component analysis remains unaffected, no matter if the data is centered or uncentered, provided that eigenvector filtering is handled appropriately.
\end{abstract}


\section{Introduction}

The covariance matrix is a primary tool in data analysis and machine learning, being the backbone of a wide range of algorithms that exploit its ability to capture linear relationships between features. Dimensionality reduction techniques like Principal Component Analysis (PCA) and Linear Discriminant Analysis (LDA) rely on the covariance matrix to identify directions of maximum variance or class separability~\cite{bishop2006pattern}. It is also crucial to density estimation in Gaussian Mixture Models (GMMs) and multivariate distributions, as well as regression and classification methods like multivariate linear regression and Quadratic Discriminant Analysis (QDA). In spectral methods such as Independent Component Analysis (ICA) and Spectral Clustering, the covariance matrix supports data preprocessing and transformation. 

Quantum computing is set to become a transformative technology, offering the potential to revolutionize fields as diverse as remote sensing. Quantum-enhanced algorithms are anticipated to surpass classical methods in efficiency and capability, particularly in the analysis of complex, high-dimensional datasets, like hyperspectral data. Properly prepared quantum covariance matrices can serve as a foundation for numerous near-term hybrid quantum-classical algorithms, such as VQSD \cite{larose2019variational} and VQSE \cite{cerezo2022variational}, and fully quantum fault-tolerant approaches, e.g.,~Quantum Principal Component Analysis \cite{lloyd2014quantum}.

\section{Amplitude encoding and the $L^2$ norm}
Amplitude encoding \cite{grover2000synthesis, plesch2011quantum, marin2023quantum} is a method in quantum computing where the components of a classical data vector are mapped to the amplitudes of a quantum state, enabling efficient representation of highly-dimensional data.
Assume one performs a quantum state preparation via an amplitude encoding
\begin{equation}
    |x_i \rangle = \frac{1}{\mathcal{N}_i}\sum_{j=0}^{n-1} x_i^j |j\rangle,
\end{equation}
where $x_i^j$ is the $j^{th}$ component of the $i^{th}$ datapoint in the dataset $\mathcal{X} = \{x_i\}_{i=1}^{m}$ and $\mathcal{N}_i$ is its $L^2$ norm, $\mathcal{N}_i = \sqrt{\sum_j(x_i^j)^2}$.
The normalization is an inherent step in amplitude encoding which assures that we retain the probabilistic interpretation of the quantum measurement.

\section{Quantum Covariance Matrix}

 As it was found in \cite{gordon2022covariance}, the average density matrix in the quantum representation of the data points is closely related to the covariance matrix for $\mathcal{X}$, $Q$,
 \begin{equation}\label{eq:Q_rho}
     Q_{ij} = \mathbb{E}\left[x^i x^j\right] - \mathbb{E}[x^i]\mathbb{E}[x^j] = \Bar{\rho} - M
 \end{equation}
where $\Bar{\rho} = \mathbb{E}\left[x^i x^j\right]$ is is the average, amplitude-encoded, density matrix, called from this point on quantum covariance matrix, and $M = \mu \otimes \mu$ is the outer product of the mean vector in $\mathcal{X}$. In the case where the data are centered ($\mu = 0$) the average density matrix is equivalent to the covariance matrix in the dataset. We identify $\mathbb{E}[\cdot]$ as an average over the data points.

\section{Standardization vs. $L^2$ norm}
Data centering and standardization are essential preprocessing steps in many data analysis methods that use the covariance matrix. 
Centering involves subtracting the mean of each feature from the dataset so that each feature has a mean of zero. 
Below we define partial centering,
\begin{equation}\label{eq:centering}
    x_i^j \mapsto x_i^j - \gamma \mu^j,
\end{equation}
where $\mu^j$ is the mean of the feature $j$ in the dataset. The parameter $\gamma\in[0,1]$ rules the strength of the centering, for $\gamma=0$ we have no centering, for $\gamma=1$ we have a full centering, and the mean of the centered data is zero in each feature.
The full centering ensures that the covariance matrix reflects only the relationships between features, rather than being skewed by feature means. 

Standardization scales each feature by its standard deviation, resulting in unit variance across features,
\begin{equation}
    x_i^j \mapsto \frac{x_i^j}{\sigma_j}.
\end{equation}
This step is crucial when features have different units or scales, as methods that rely on the covariance matrix are sensitive to the relative magnitudes of feature variances. 
Without centering and standardization, these methods may produce biased results or fail to capture the true structure of the data.

\begin{figure}[ht!]
    \centering
    \includegraphics[width=0.7\linewidth]{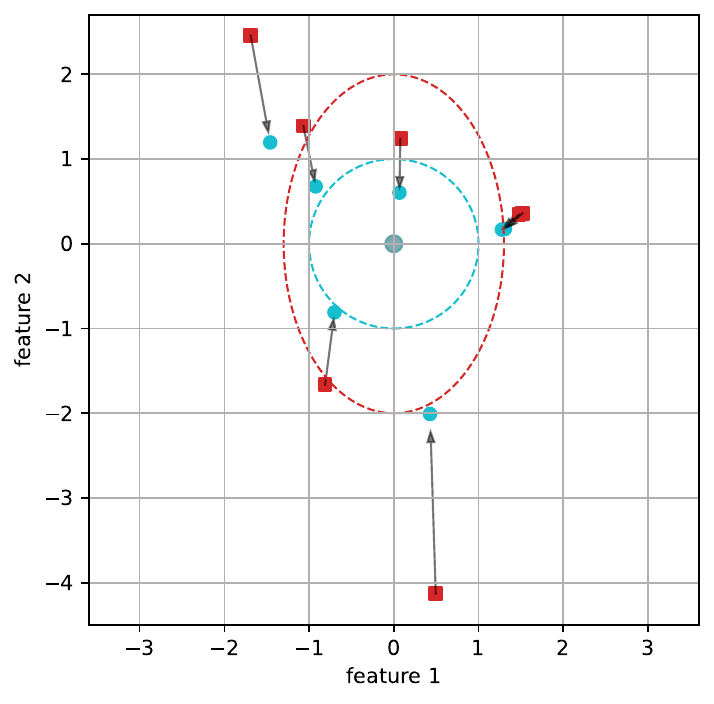}
    \includegraphics[width=0.7\linewidth]{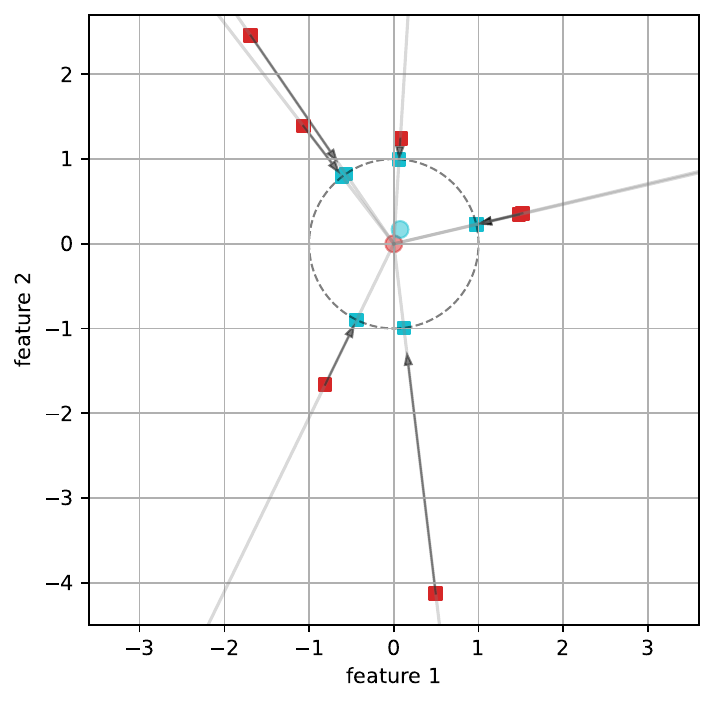}
    \caption{Comparison of the data standardization and $L^2$-normalization of the initially centered data. Blue and orange squares represent data points before and after the transformation. The semi-transparent blue and orange circles indicate the mean feature values, respectively, before and after the transformation.  \textit{Top:} Data standardization consists of the rescaling of each feature $j$ by the inverse its standard deviation~$\sigma^j$. Each feature of the data point is rescaled proportionally with respect to the symmetry around $0$ in this direction. Thus the centering is not affected by the transformation. \textit{Bottom:} $L^2$-normalization consists of the radial projection of the data points on the unit sphere. Each vector is normalized separately, and the final position of the point does not depend on its magnitude but rather on its angle with respect to the coordinate system used. If the initial points do not possess spherical symmetry, the data centering is affected by the $L^2$-normalization.}
    \label{fig:data_processing}
\end{figure}

For the amplitude-encoded quantum data, achieving correct centering of the data is in general impossible (the only exception is when the data has a spherical symmetry), as the last step, before encoding, of the data processing is to $L^2$-normalize the data. 
Contrary to standardization, $L^2$-normalization does influence centering. 
The schematic representation of standardization and $L^2$ normalization can be seen in Fig. \ref{fig:data_processing}. The figure shows the idea behind the influence of the $L^2$-normalization on data centering. 

It has been observed that the PCA based on centered and uncentered data can have surprisingly a lot in common \cite{cadima2009relationships}.
Thus, we ask the following research question in this work: \textbf{How does the incompatibility of centering and $L^2$ normalization influence the performance of the data analysis methods based on the covariance matrices?}

\begin{figure*}[ht!]
    \centering
    \includegraphics[width=0.85\linewidth]{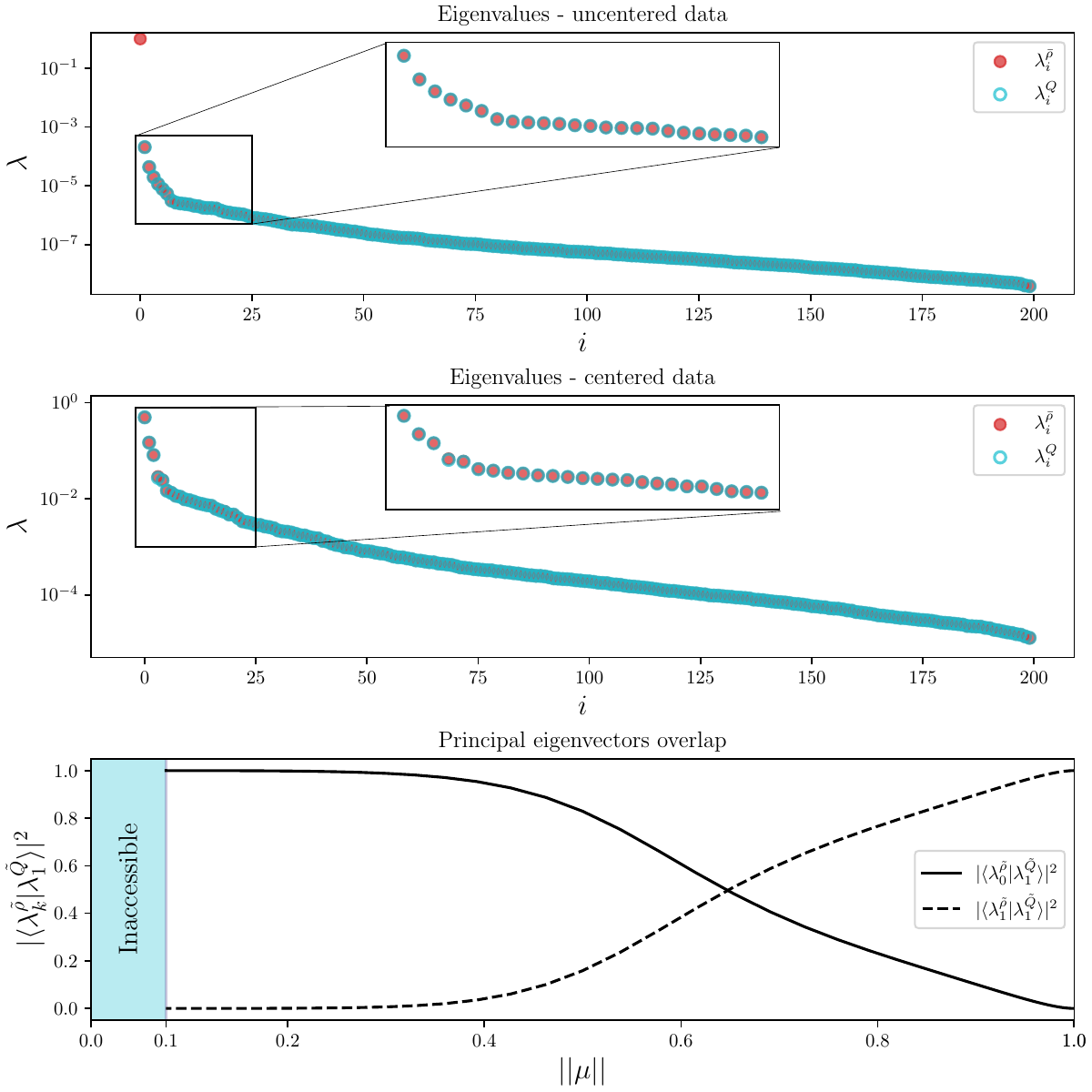}
    \caption{The impact of the data centering on the spectrum and the eigenvectors of quantum covariance matrix. \textit{Top:} Eigenvalues of the covariance matrix $\lambda^Q$'s and the quantum covariance matrix $\lambda^{\Bar{\rho}}$'s for the uncentered data. The first eigenvalue of $\Bar{\rho}$ is related to the data mean $\mu$ and does not have a counterpart in the spectrum of $Q$. Shifting the eigenvalues by one of the compared spectra shows almost perfect equivalence between remaining eigenvalues. \textit{Middle:} Eigenvalues of the covariance matrix $\lambda^Q$'s and the quantum covariance matrix $\lambda^{\Bar{\rho}}$'s for the centered data. The spectra of $Q$ and $\bar
    \rho$ match closely. \textit{Bottom}: Fidelity between the first eigenvector of the covariance matrix $\lambda^Q_1$ and the first two eigenvectors of the quantum covariance matrix $\lambda^{\Bar{\rho}}_{0/1}$ as a function of the norm of the mean vector $\mu$ after $L^2$ normalization. The crossing value of $||\mu||\approx 0.65$ corresponds to the centering parameter $\gamma\approx0.98$ in Eq.~(\ref{eq:centering}). The inaccessible region marks the $||\mu||$ values which were not accessible by a single centering followed by $L^2$ normalization.}
    \label{fig:eigenanalysis}
\end{figure*}

\section{Eigenproblem for $Q$ and $\bar{\rho}$}
In this study, we perform the following data preprocessing:
$$\text{Standardization} \mapsto \text{(partial) centering} \mapsto L^2\ \text{normalization}$$
We begin with the analysis of the spectra and eigenvectors of the two matrices $Q$ and $\bar{\rho}$ for different centerings. 
In Fig.~\ref{fig:eigenanalysis}, we analyze points labelled with two classes (\textit{Corn-mintill}, \textit{Soybean-notill}) for the \texttt{Indian Pines} hyperspectral dataset~\cite{baumgardner220BandAviris2015}. Hyperspectral images capture a large number of contiguous spectral bands which might reveal intrinsic characteristics of the scanned objects. 
For the uncentered data ($\gamma=0$ in Eq.~\ref{eq:centering}) we obtain a principal eigenvalue in $\bar{\rho}$, which does not have counterpart in the $Q$ spectrum. It corresponds to the mean feature values vector after $L^2$ normalization, $\mu$. As all other eigenvalues between mentioned matrices match closely, we conclude that the $\mu$ vector is orthogonal to the eigenspace of $Q$.
In the case of the centered data, we still obtain non-zero $\mu$ vector after the $L^2$ normalization, however it seems that the resulting non-centering does not significantly influence the spectrum of $\bar{\rho}$, when compared with the spectrum of $Q$.

\begin{table*}[ht!]
    \centering
    \caption{The RBF-kernel SVM-based binary classification mean accuracy (standard deviation of accuracy) for the classes: 3/10; 2/11; 5/8 for the \texttt{Indian Pines} dataset \cite{baumgardner220BandAviris2015}. The results are obtained for a 5-fold cross-validation. Different preprocessing schemes are denoted as follows: CL - classical preprocessing, standarization, centering, no $L^2$ normalization, PCA based on $Q$; UC - standardization, uncentered data, PCA based on $\bar{\rho}$; UC-skip - standardization, uncentered data, PCA based on $\bar{\rho}$ with the omission of the first eigenvector; C - standardization, centered data, PCA based on $\bar{\rho}$; HC - standardization, partially centered data ($\gamma=0.95$), PCA based on $\bar{\rho}$. $n$ indicates the number of principal components taken for the classification.}
    \label{tab:acc_table}
    \resizebox{1.9\columnwidth}{!}{%
\begin{tabular}{ll|ll|ll|ll|ll|ll|}
\toprule
    &      & \multicolumn{2}{c}{CL} & \multicolumn{2}{c}{UC} & \multicolumn{2}{c}{UC-skip} & \multicolumn{2}{c}{C} & \multicolumn{2}{c}{HC} \\
    &      &       Train &        Test &       Train &        Test &       Train &        Test &       Train &        Test &       Train &        Test \\
Task & n &             &             &             &             &             &             &             &             &             &             \\
\midrule
3/10 & 2.0  &  0.86(0.01) &  0.85(0.03) &  0.54(0.01) &  0.54(0.02) &  0.86(0.00) &  0.86(0.03) &  0.85(0.00) &  0.84(0.02) &  0.60(0.01) &  0.59(0.04) \\
    & 3.0  &  0.92(0.00) &  0.91(0.02) &  0.54(0.01) &  0.54(0.02) &  0.93(0.00) &  0.93(0.02) &  0.92(0.01) &  0.92(0.02) &  0.81(0.01) &  0.80(0.03) \\
    & 4.0  &  0.97(0.00) &  0.96(0.01) &  0.54(0.01) &  0.54(0.02) &  0.97(0.00) &  0.97(0.00) &  0.98(0.00) &  0.98(0.01) &  0.84(0.01) &  0.83(0.01) \\
    & 5.0  &  0.98(0.00) &  0.98(0.00) &  0.54(0.01) &  0.54(0.02) &  0.99(0.00) &  0.99(0.00) &  0.99(0.00) &  0.99(0.00) &  0.94(0.00) &  0.94(0.01) \\
    & 10.0 &  0.99(0.00) &  0.98(0.01) &  0.54(0.01) &  0.54(0.02) &  0.99(0.00) &  0.99(0.01) &  0.99(0.00) &  0.99(0.01) &  0.94(0.00) &  0.94(0.01) \\ \midrule[1pt]
2/11 & 2.0  &  0.78(0.00) &  0.77(0.01) &  0.63(0.00) &  0.63(0.01) &  0.78(0.00) &  0.77(0.01) &  0.75(0.01) &  0.74(0.01) &  0.72(0.00) &  0.72(0.01) \\
    & 3.0  &  0.78(0.00) &  0.78(0.01) &  0.63(0.00) &  0.63(0.01) &  0.78(0.00) &  0.78(0.01) &  0.77(0.01) &  0.76(0.02) &  0.77(0.00) &  0.77(0.01) \\
    & 4.0  &  0.79(0.00) &  0.78(0.01) &  0.63(0.00) &  0.63(0.01) &  0.79(0.00) &  0.79(0.01) &  0.82(0.00) &  0.81(0.01) &  0.76(0.00) &  0.76(0.01) \\
    & 5.0  &  0.81(0.01) &  0.81(0.01) &  0.63(0.00) &  0.63(0.01) &  0.82(0.01) &  0.81(0.01) &  0.84(0.01) &  0.83(0.01) &  0.77(0.00) &  0.77(0.02) \\
    & 10.0 &  0.89(0.00) &  0.88(0.01) &  0.63(0.00) &  0.63(0.01) &  0.90(0.00) &  0.89(0.01) &  0.89(0.00) &  0.88(0.01) &  0.85(0.00) &  0.84(0.01) \\ \midrule[1pt]
5/8 & 2.0  &  0.96(0.00) &  0.96(0.01) &  0.83(0.01) &  0.83(0.03) &  0.96(0.00) &  0.96(0.01) &  0.96(0.00) &  0.96(0.01) &  0.88(0.00) &  0.88(0.02) \\
    & 3.0  &  0.97(0.00) &  0.97(0.01) &  0.83(0.01) &  0.83(0.03) &  0.98(0.00) &  0.98(0.01) &  0.97(0.00) &  0.97(0.01) &  0.96(0.00) &  0.96(0.01) \\
    & 4.0  &  0.98(0.00) &  0.98(0.01) &  0.83(0.01) &  0.83(0.02) &  0.98(0.00) &  0.98(0.01) &  0.98(0.00) &  0.98(0.01) &  0.96(0.00) &  0.96(0.01) \\
    & 5.0  &  0.99(0.00) &  0.99(0.01) &  0.83(0.01) &  0.83(0.02) &  0.99(0.00) &  0.99(0.00) &  0.98(0.00) &  0.98(0.01) &  0.97(0.00) &  0.96(0.01) \\
    & 10.0 &  0.99(0.00) &  0.99(0.01) &  0.83(0.01) &  0.83(0.02) &  1.00(0.00) &  0.99(0.00) &  0.99(0.00) &  0.99(0.01) &  0.99(0.00) &  0.98(0.01) \\
\bottomrule
\end{tabular}
}
\end{table*}

In the bottom plot of Fig.~\ref{fig:eigenanalysis} we study how does the first eigenvector of $Q$ is related to the two first eigenvectors of $\Bar{\rho}$ as a function of the---after the $L^2$ normalization---mean feature vector norm $||\mu||$.
For high $||\mu||$ values (highly uncentered dataset) it's the second eigenvector of $\Bar{\rho}$ which has a notable overlap with the first eigenvector of $Q$. We conjecture that in such case, the first eigenvector of $\Bar{\rho}$ should be dropped from the analysis as it represents the $\mu$, while not influencing the eigen-subspace of $\Bar{\rho}$ corresponding to the eigenspace of $Q$. For almost-centered datasets (low $||\mu||$ values), the resulting mean vector $\mu$ is not clearly manifested in the eigenproblem of $\bar{\rho}$, the $M$ matrix (Eq.~\ref{eq:Q_rho}) does not have a significant influence, rendering $Q$ and $\bar{\rho}$ almost equivalent. For almost centered datasets, all eigenvectors of $\Bar{\rho}$ should be taken into account for the covariance-matrix-based analysis.
One can note that the correspondence between partial centering parameter $\gamma$ and the value of $||\mu||$ is non-linear. The crossing point on the bottom plot is for $||\mu|| \approx 0.65$, which corresponds to almost centered dataset, $\gamma=0.98$. Therefore, the $||\mu||$ value can be highly sensitive for any non-centering in the dataset. If we choose to first center the dataset, and then use full $\Bar{\rho}$ as the covariance matrix, we risk ending up in the regime where the principal eigenvector of $Q$ is distributed between first two eigenvectors of $\Bar{\rho}$. This in turn, would inevitably lead to poor performance in the covariance-matrix-based data analysis task.

\section{Impact on the classification}

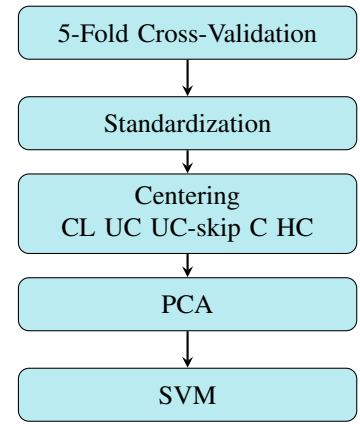
\begin{figure}[ht!]
    \begin{center}
\begin{tikzpicture}[node distance=1.2cm]

\node (crossvalidation) [block] {5-Fold Cross-Validation};
\node (standardization) [block, below of=crossvalidation] {Standardization};
\node (centering) [doubleblock, below of=standardization] {Centering \\ CL UC UC-skip C HC};
\node (pca) [block, below of=centering] {PCA};
\node (svm) [block, below of=pca] {SVM};

\draw [arrow] (crossvalidation) -- (standardization);
\draw [arrow] (standardization) -- (centering);
\draw [arrow] (centering) -- (pca);
\draw [arrow] (pca) -- (svm);

\end{tikzpicture}
\end{center}
    \caption{The high-level analysis flowchart to verify the impact of the data preprocessing on the classification performance.}
    \label{fig:analysis_flowchart}
\end{figure}

Now, we focus on the impact of the data preprocessing on the classification accuracy in the PCA-reduced dataset. We take the points labeled with the classes 2---\textit{Corn-notill}; 3---\textit{Corn-mintill}; 5---\textit{Grass-pasture}; 8---\textit{Hay-windrowed}; 10---\textit{Soybean-notill}; 11---\textit{Soybean-mintill}, produce 5-fold cross validation, apply PCA dimensionality reduction and perform binary classification for the pairs: 2/11; 3/10; 5/8. The classifier employed is an SVM with the radial basis function kernel, and the workflow of the analysis process is rendered in Fig.~\ref{fig:analysis_flowchart}.  The centering methods are explained in the caption of Table \ref{tab:acc_table}.


The classification results are presented in Table \ref{tab:acc_table}.
They indicate that choosing a proper preprocessing model does matter. When compared with the classical preprocessing (CL), the UC-skip and C models show a competetive performance. This agrees with the intuition that quantum covariance matrices $\Bar{\rho}$ in both of those models have a similar spectrum as the classical covariance matrix $Q$. The model UC includes an eigenvector connected to the centering, which sabotages the classification, while in the HC model, this eigenvector is partially present in both first and second eigenvectors of $\Bar{\rho}$, which has a detrimental influence on the classification performance.

\section{Conclusions}
In the quantum covariance matrix preparation the hyperspectral data preprocessing does matter. For a reliable covariance matrix-based data analysis one of two approaches are advised: a) fully center the data before amplitude encoding, b) do not center the data, but drop the principal eigenvector of the quantum covariance matrix from the further analysis. Although both methods gave a satisfying results in the studied dataset, the b) approach seems more robust, as with the approach a) one can potentially easily end up with uncentered data after $L^2$ normalization, which leads to the inferior HC strategy.

\small
\bibliographystyle{IEEEtranN}
\bibliography{references}

\begin{thebibliography}{10}
\providecommand{\natexlab}[1]{#1}
\providecommand{\url}[1]{#1}
\csname url@samestyle\endcsname
\providecommand{\newblock}{\relax}
\providecommand{\bibinfo}[2]{#2}
\providecommand{\BIBentrySTDinterwordspacing}{\spaceskip=0pt\relax}
\providecommand{\BIBentryALTinterwordstretchfactor}{4}
\providecommand{\BIBentryALTinterwordspacing}{\spaceskip=\fontdimen2\font plus
\BIBentryALTinterwordstretchfactor\fontdimen3\font minus \fontdimen4\font\relax}
\providecommand{\BIBforeignlanguage}[2]{{%
\expandafter\ifx\csname l@#1\endcsname\relax
\typeout{** WARNING: IEEEtranN.bst: No hyphenation pattern has been}%
\typeout{** loaded for the language `#1'. Using the pattern for}%
\typeout{** the default language instead.}%
\else
\language=\csname l@#1\endcsname
\fi
#2}}
\providecommand{\BIBdecl}{\relax}
\BIBdecl

\bibitem[Bishop and Nasrabadi(2006)]{bishop2006pattern}
C.~M. Bishop and N.~M. Nasrabadi, \emph{Pattern recognition and machine learning}.\hskip 1em plus 0.5em minus 0.4em\relax Springer, 2006, vol.~4, no.~4.

\bibitem[LaRose et~al.(2019)LaRose, Tikku, O’Neel-Judy, Cincio, and Coles]{larose2019variational}
R.~LaRose, A.~Tikku, {\'E}.~O’Neel-Judy, L.~Cincio, and P.~J. Coles, ``Variational quantum state diagonalization,'' \emph{npj Quantum Information}, vol.~5, no.~1, p.~57, 2019.

\bibitem[Cerezo et~al.(2022)Cerezo, Sharma, Arrasmith, and Coles]{cerezo2022variational}
M.~Cerezo, K.~Sharma, A.~Arrasmith, and P.~J. Coles, ``Variational quantum state eigensolver,'' \emph{npj Quantum Information}, vol.~8, no.~1, p. 113, 2022.

\bibitem[Lloyd et~al.(2014)Lloyd, Mohseni, and Rebentrost]{lloyd2014quantum}
S.~Lloyd, M.~Mohseni, and P.~Rebentrost, ``Quantum principal component analysis,'' \emph{Nature physics}, vol.~10, no.~9, pp. 631--633, 2014.

\bibitem[Grover(2000)]{grover2000synthesis}
L.~K. Grover, ``Synthesis of quantum superpositions by quantum computation,'' \emph{Physical review letters}, vol.~85, no.~6, p. 1334, 2000.

\bibitem[Plesch and Brukner(2011)]{plesch2011quantum}
M.~Plesch and {\v{C}}.~Brukner, ``Quantum-state preparation with universal gate decompositions,'' \emph{Physical Review A—Atomic, Molecular, and Optical Physics}, vol.~83, no.~3, p. 032302, 2011.

\bibitem[Marin-Sanchez et~al.(2023)Marin-Sanchez, Gonzalez-Conde, and Sanz]{marin2023quantum}
G.~Marin-Sanchez, J.~Gonzalez-Conde, and M.~Sanz, ``Quantum algorithms for approximate function loading,'' \emph{Physical Review Research}, vol.~5, no.~3, p. 033114, 2023.

\bibitem[Gordon et~al.(2022)Gordon, Cerezo, Cincio, and Coles]{gordon2022covariance}
M.~H. Gordon, M.~Cerezo, L.~Cincio, and P.~J. Coles, ``Covariance matrix preparation for quantum principal component analysis,'' \emph{PRX Quantum}, vol.~3, no.~3, p. 030334, 2022.

\bibitem[Cadima and Jolliffe(2009)]{cadima2009relationships}
J.~Cadima and I.~Jolliffe, ``On relationships between uncentred and column-centred principal component analysis.'' \emph{Pakistan Journal of Statistics}, vol.~25, no.~4, 2009.

\bibitem[Baumgardner et~al.(2015)Baumgardner, Biehl, and Landgrebe]{baumgardner220BandAviris2015}
M.~F. Baumgardner, L.~L. Biehl, and D.~A. Landgrebe, ``220 band aviris hyperspectral image data set: {{June}} 12, 1992 indian pine test site 3,'' \emph{Purdue University Research Repository}, vol.~10, no.~7, p. 991, 2015.

\end{thebibliography}

\end{document}